\documentclass[twocolumn]{aastex631}

\usepackage{graphicx}	% Including figure files
\usepackage{amsmath}	% Advanced maths commands

\begin{document}

\title[]{Periodic variable star classification with deep learning: \\ handling data imbalance in an ensemble augmentation way}

\author[0000-0003-3778-3566]{Zihan Kang}
\affiliation{Key Laboratory of Optical Astronomy, National Astronomical Observatories, Chinese Academy of Sciences, Beijing 100012, China}
\affiliation{National Astronomical Observatories, Chinese Academy of Sciences, Beijing 100012, China}
\affiliation{University of Chinese Academy of Science, Beijing 100049, China}

\author{Yanxia Zhang}\thanks{E-mail: zyx@bao.ac.cn}
\affiliation{Key Laboratory of Optical Astronomy, National Astronomical Observatories, Chinese Academy of Sciences, Beijing 100012, China}
\affiliation{National Astronomical Observatories, Chinese Academy of Sciences, Beijing 100012, China}

\author{Jingyi Zhang}\thanks{E-mail: jyzhang@bao.ac.cn}
\affiliation{Key Laboratory of Optical Astronomy, National Astronomical Observatories, Chinese Academy of Sciences, Beijing 100012, China}
\affiliation{National Astronomical Observatories, Chinese Academy of Sciences, Beijing 100012, China}

\author{Changhua Li}
\affiliation{Key Laboratory of Optical Astronomy, National Astronomical Observatories, Chinese Academy of Sciences, Beijing 100012, China}

\author{Minzhi Kong}
\affiliation{College of Physics, Hebei Normal University, Shijiazhuang 050024, China}

\author{Yongheng Zhao}
\affiliation{Key Laboratory of Optical Astronomy, National Astronomical Observatories, Chinese Academy of Sciences, Beijing 100012, China}
\affiliation{National Astronomical Observatories, Chinese Academy of Sciences, Beijing 100012, China}

\author{Xue-Bing Wu}
\affiliation{Department of Astronomy, School of Physics, Peking University, Beijing 100871, China}
\affiliation{Kavli Institute for Astronomy and Astrophysics, Peking University, Beijing 100871, China}

\begin{abstract}
Time-domain astronomy is progressing rapidly with the ongoing and upcoming large-scale photometric sky surveys led by the Vera C. Rubin Observatory project (LSST). Billions of variable sources call for better automatic classification algorithms for light curves. Among them, periodic variable stars are frequently studied. Different categories of periodic variable stars have a high degree of class imbalance and pose a challenge to algorithms including deep learning methods. We design two kinds of architectures of neural networks for the classification of periodic variable stars in the Catalina Survey's Data Release 2: a multi-input recurrent neural network (RNN) and a compound network combing the RNN and the convolutional neural network (CNN). To deal with class imbalance, we apply Gaussian Process to generate synthetic light curves with artificial uncertainties for data augmentation. For better performance, we organize the augmentation and training process in a ``bagging-like" ensemble learning scheme. The experimental results show that the better approach is the compound network combing RNN and CNN, which reaches the best result of 86.2 per~cent on the overall balanced accuracy and $0.75$ on the macro F1 score. We develop the ensemble augmentation method to solve the data imbalance when classifying variable stars and prove the effectiveness of combining different representations of light curves in a single model. The proposed methods would help build better classification algorithms of periodic time series data for future sky surveys (e.g. LSST).
\end{abstract}


\section{Introduction}

With the upcoming large-scale sky surveys represented by the Vera C. Rubin Observatory project \citep[LSST;][]{2019ApJ...873..111I}, time-domain astronomy is now entering a golden age with overwhelming data. Besides, the ongoing surveys are still accumulating data to be analyzed, such as the Zwicky Transient Facility \citep[ZTF;][]{2019PASP..131a8002B}, the Optical Gravitational Lensing Experiment \citep[OGLE;][]{udalski2015ogleiv}, and the Catalina Real-Time Transient Survey \citep[CRTS;][]{2009ApJ...696..870D}. Billions of observed variable sources demand automatic classification for further research. Machine learning plays a prominent role in this task and is broadly divided into two types of algorithms: traditional approaches with artificial input features and deep learning methods based on various neural networks. 

Among numerous kinds of variable sources, periodic variable stars are often studied as a particular category because of their importance and distinct observable difference from other sources. For example, Cepheid stars and RR Lyrae stars can be used as standard candles for distance measurement \citep[][]{2003LNP...635.....A}, hence are crucial for the Galaxy structure. However, the samples of different periodic variable star classes are highly imbalanced, meaning some classes dominate the known samples while others have few cases. This makes it challenging to train a satisfying machine learning model for the classification task due to the scarcity of some samples and the bias towards the majority classes.

Techniques for dealing with the class imbalance in machine learning can be grouped into three categories: data-level, algorithm-level, and hybrid approaches \citep[][]{henning2023survey}. Data-level methods focus on adjusting the training data by resampling and augmentation. Algorithm-level techniques generally modify algorithms with a weight or cost schema, with the assumption that the data are sufficient. Hybrid approaches combine both of them and are often implemented with ensemble learning methods. For traditional machine learning algorithms with artificial input features, plenty of researches are aimed at imbalance learning, such as the Synthetic Minority Over-sampling Technique \citep[SMOTE;][]{Chawla_2002} and the Self-paced Ensemble \citep[SPE;][]{Liu_2020}. Nevertheless, the subject of deep learning with class-imbalanced data is understudied \citep[][]{henning2023survey}. Although some techniques exist for deep learning to avoid bias towards majority classes on the imbalance data, the issue remains due to the lack of data. Deep learning relies much more on data sufficiency than traditional machine learning algorithms because of the high model complicacy with many parameters. A practical solution would be to find an ideal data augmentation method, especially for astronomical light curves. Physical models and data-driven methods are often applied to generate synthetic light curves for data augmentation. As for variable stars, there are no proper physical models for now, only data-driven approaches to be considered, including adding noise \citep[][]{2018NatAs...2..151N}, the Gaussian Process \citep[][]{Faraway_2016, Castro_2017} and the deep generative models \citep[][]{Palomera_2022}. 

The recurrent neural network (RNN) is a widely used deep learning algorithm for light curves, accepting both the light curve and its uncertainties as input \citep[][]{2018NatAs...2..151N}. RNN can also combine with scalar contextual information, such as period and colour, to form a multi-input model \citep[][]{2021MNRAS.505.4345B}. A synthetic light curve for an RNN model should have a well-modelled uncertainty, which is ignored by previous studies. There is another deep learning method to classify variable stars using the convolutional neural network (CNN) presented by \citet{2022ApJ...938...37S}. CNN takes the light curve image as input, which also demands Gaussian Process augmentation.

Our work aims to find suitable neural network architecture for periodic variable star classification on the CRTS data while applying proper augmentation to handle data imbalance. We design a multi-input RNN-based neural network and apply Gaussian Process to generate artificial light curves with uncertainties. We also develop a compound neural network architecture by combining the RNN and CNN structures. To mitigate model overfitting on minority classes while improving classification performance, we organize the augmentation and training process in a ``bagging-like" ensemble learning scheme. 

This paper is organized as follows. Section \ref{Data And Augmentation} briefly introduces the CRTS variable star data and details the Gaussian process augmentation. Section \ref{Ensemble Learning Method} describes the ensemble learning scheme we adopt. In Section \ref{Multi-input Neural Networks}, we characterize the two neural network architectures we design. Section \ref{implementation and results} gives the classification result by a comprehensive evaluation on an imbalance test data set. Section \ref{sec:discussion} presents some limitations and future work, and finally, the summary is provided in Section \ref{Conclusions}.  We publish our source code on \url{https://github.com/52Hzihan/mixnn4vs}.

\section{Data And Augmentation}
\label{Data And Augmentation}
The Catalina Real-Time Transient Survey (CRTS) surveyed 33,000 deg$^2$ of the sky and produced more than 500 million light curves for various sources. Among them, CRTS DR2 provided a catalogue for variable stars \citep[][]{2017MNRAS.469.3688D}. Similar to \citet[][]{2020MNRAS.493.6050H}, we take 11 classes into account for our analysis, as presented in Table~\ref{tab:data entries}. In addition, CRTS uses an unfiltered telescope, so the lack of colour data highlights the importance of extracting information from light curves when implementing classification.

\begin{table}
	\centering
	\caption{The number of different classes of CRTS variable stars.}
	\label{tab:data entries}
	\begin{tabular}{l|l} % four columns, alignment for each
		\hline
		Classes of variable stars & No. \\
        \hline
        RRab & 4325 \\
        Blazhko & 171\\
        RRc & 3752 \\
        RRd & 502 \\
        Rot (Rotational) & 3636 \\
        Ecl (Contact and Semi-Detached Binary) & 18803 \\
        EA (Detached Binary) & 4509 \\
        LPV (Long Period Variable) & 1286 \\
        $\delta$-Scuti & 147 \\
        ACEP (Anomalous Cepheids) & 153 \\
        Cep-\uppercase\expandafter{\romannumeral2} (Type-\uppercase\expandafter{\romannumeral2} Cepheids) & 153\\

		\hline
	\end{tabular}
\end{table}

To clean the dataset, we fit the phase-folded light curves with Friedman's SuperSmoother \citep[][]{Friedman1984AVS} and exclude data points deviating more than three standard deviations from the smoothed light curves. We also delete points with significant errors greater than twice the average error.

\subsection{Gaussian Process}
In order to deal with the imbalance of data, we need to generate simulated light curves for augmentation. Since no proper physical model is available for all types of variable stars,  the only way is to create synthetic data from natural light curves. Because the deep learning approach we adopt takes uncertainties as part of inputs, we ought to build models for both natural light curves and their uncertainties. Therefore we turn to Gaussian Process (GP; \citealt{2006gpml.book.....R}), a stochastic process for modelling time series, which is applied for light curve data augmentation in previous studies \citep[][]{2019AJ....158..257B}.

GP is a distribution over functions. In our simple form with scalar input, it can be viewed as an Infinite-dimensional joint Gaussian distribution over time, which is fully described by its mean function and kernel function (i.e. covariance function).

\begin{equation}
    f(t) \sim GP(\mu(t), k(t,t'))
\end{equation}
where $\mu(t)$ is the mean function and $k(t,t')$ computes the covariance between two points $t$ and $t'$.

To fit a GP model for a light curve, we need to choose a prior kernel function and a prior mean function, then calculate the Bayesian posterior functions under the data. Here we adopt the Matern $5/2$ kernel as the prior kernel, which is given by

\begin{equation}
    k_{Matern52}(\tau)=\alpha^2\left(1+\frac{\sqrt{5}\tau}{\rho}+\frac{5\tau^2}{3\rho^2}\right)\exp\left(-\frac{\sqrt{5}\tau}{\rho}\right)
\end{equation}
where $\tau=t-t'$, $\alpha$ and $\rho$ are the hyperparameters to be optimized. For the prior mean function, we can simply set $\mu(t)=0$.

Given a real light curve with uncertainties $( \textbf{t}, \textbf{m},\pmb{\sigma})$, For any other random time points $\textbf{t}_*$, the joint distribution of $\textbf{m}$ and the predicted magnitude $\textbf{m}_*$ will be a multivariate Gaussian distribution as follow:

\begin{equation}
    \genfrac{[}{]}{0pt}{}{\textbf{m}}{\textbf{m}_*} \sim \mathcal{N}\left(\genfrac{[}{]}{0pt}{}{\mu(\textbf{t})}{\mu(\textbf{t}_*)}, \begin{bmatrix}k(\textbf{t}, \textbf{t})+diag(\pmb{\sigma})^2&k(\textbf{t}, \textbf{t}_*)\\k(\textbf{t}_*, \textbf{t})&k(\textbf{t}_*, \textbf{t}_*)\end{bmatrix}\right)
\end{equation}
The posterior distribution of $\textbf{m}_*$ comes as $\textbf{m}_*\sim\mathcal{N}(\overline{\mu}(\textbf{t}_*), \overline{k}(\textbf{t}_*,\textbf{t}_*))$, where

\begin{gather}
    \overline{\mu}(\textbf{t}_*) = k(\textbf{t}_*, \textbf{t})[k(\textbf{t}, \textbf{t})+diag(\pmb{\sigma})^2]^{-1}(\textbf{m}-\mu(\textbf{t})) + \mu(\textbf{t}_*)\\
    \overline{k}(\textbf{t}_*,\textbf{t}_*) = k(\textbf{t}_*, \textbf{t}_*)-k(\textbf{t}_*, \textbf{t})[k(\textbf{t}, \textbf{t})+diag(\pmb{\sigma})^2]^{-1}k(\textbf{t}, \textbf{t}_*)
\end{gather}

The hyperparameters of the prior kernel function are optimized by minimizing the negative log-likelihood function 
\begin{equation}
    \ln\mathcal{L}(\alpha, \rho)=-\frac{1}{2}\textbf{r}^T\textbf{K}^{-1}\textbf{r}-\frac{1}{2}\ln\det(\textbf{K})-\frac{N}{2}\ln(2\pi)
\end{equation}
where $\textbf{K} = k(\textbf{t}, \textbf{t})+diag(\pmb{\sigma})^2$, $\textbf{r}$ is the residual after subtracting the model prediction means from the observations, and $N$ is the number of data points.

We employ the GP regression using \textbf{George} \citep[][]{2015ITPAM..38..252A}.

\subsection{Generate Synthetic Light Curves with Uncertainty}
\label{subsec:Generate Synthetic Light Curves}

The GP allows us to generate synthetic light curves $( \textbf{t}_*, \textbf{m}_*,\pmb{\sigma}_*)$ on randomly sampled time points $\textbf{t}_*$, where $\pmb{\sigma}_*=tr(\overline{k}(\textbf{t}_*, \textbf{t}_*))$. To make a synthetic light curve more "real" in the specific generation process, we scale up $\pmb{\sigma}_*$ to ensure the mean error is the same as its prototype. The time points $\textbf{t}_*$ are sampled to have the same size as $\textbf{t}$, and the magnitudes $\textbf{m}_*$ come from the corresponding $\overline{\mu}(\textbf{t}_*)$ adding Gaussian noises with scaled $\pmb{\sigma}_*$ as standard deviations. In addition, we apply a random phase shift for each synthetic light curve to enhance diversity.

\begin{figure*}

    \centering
    \includegraphics[width=\textwidth]{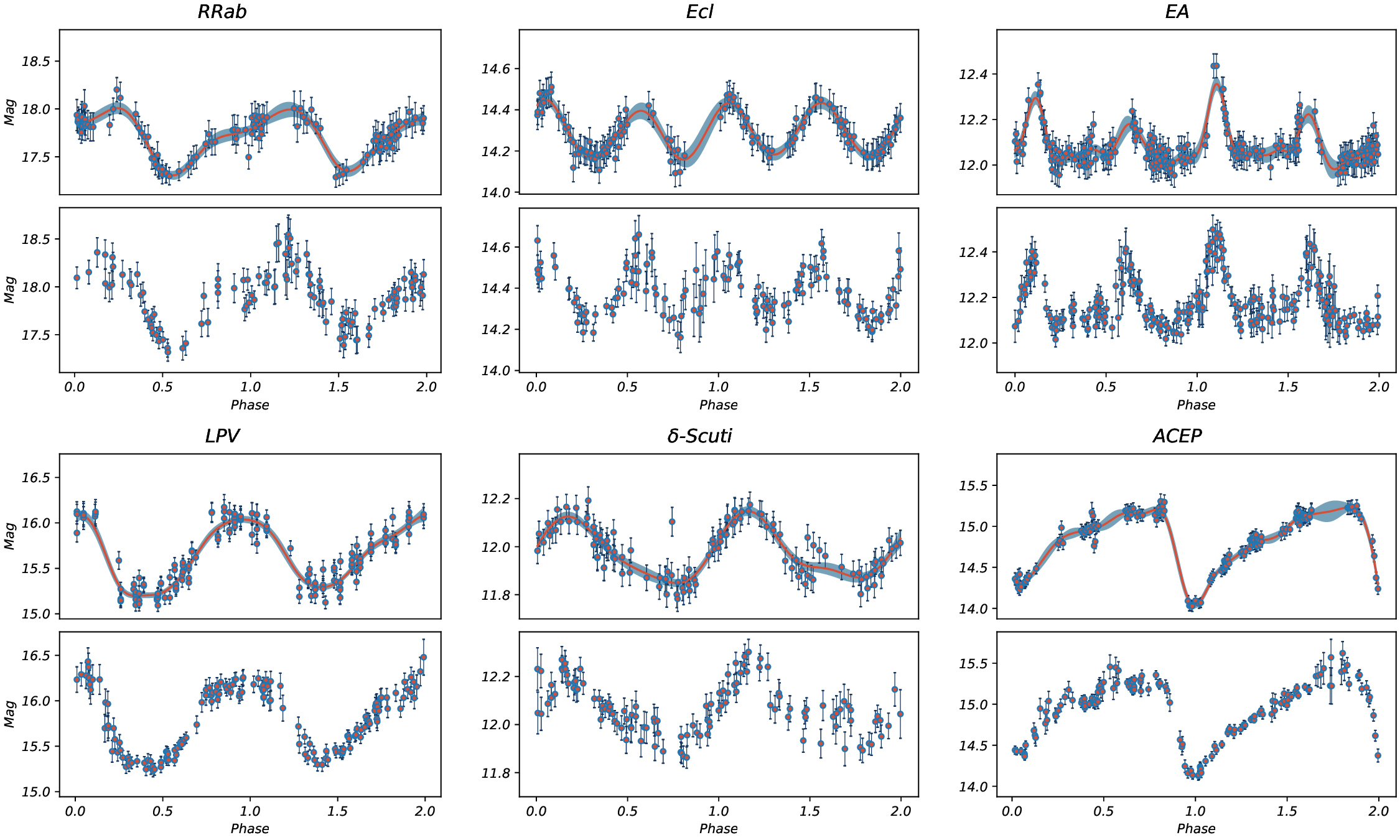}
    \caption{Examples of GP regression and synthetic light curves on several classes. For these six variable stars, the above plotted the natural light curve with error bars and its GP model, and the bottom plotted a corresponding synthetic light curve with synthetic error bars. The mean functions of the GP models are presented in solid lines, and the modelled uncertainties are illustrated in filled blue regions.}
    \label{Gaussian_Process}
\end{figure*}

Figure~\ref{Gaussian_Process} shows examples of GP regression and synthetic light curves on several classes. The light curves are folded with twice the period for better exhibition. As a data-driven model, the GP regression result gives significant uncertainty at sections with few observations.

\section{Ensemble Learning Method}
\label{Ensemble Learning Method}

A general approach for data augmentation is to produce enough synthetic light curves so that every class has an equal size for training data. However, in the case of the deep learning method, especially the recurrent neural network we adopt, this equal augmentation method meets problems. Although an artificial light curve has different numerical values against its prototype, their high-level features in the neural networks may still resemble each other since their shapes look similar. For categories with few entries, equal augmentation means too many simulations for a single light curve, which may cause overfitting on samples of small-size classes when training the neural networks. Employing fewer simulations and applying class weights may partly overcome the overfitting problem; however, this is still some trade-off between large-size and small-size classes, and it makes a limited contribution to the overall performance.

\begin{figure*}

    \includegraphics[width=\textwidth]{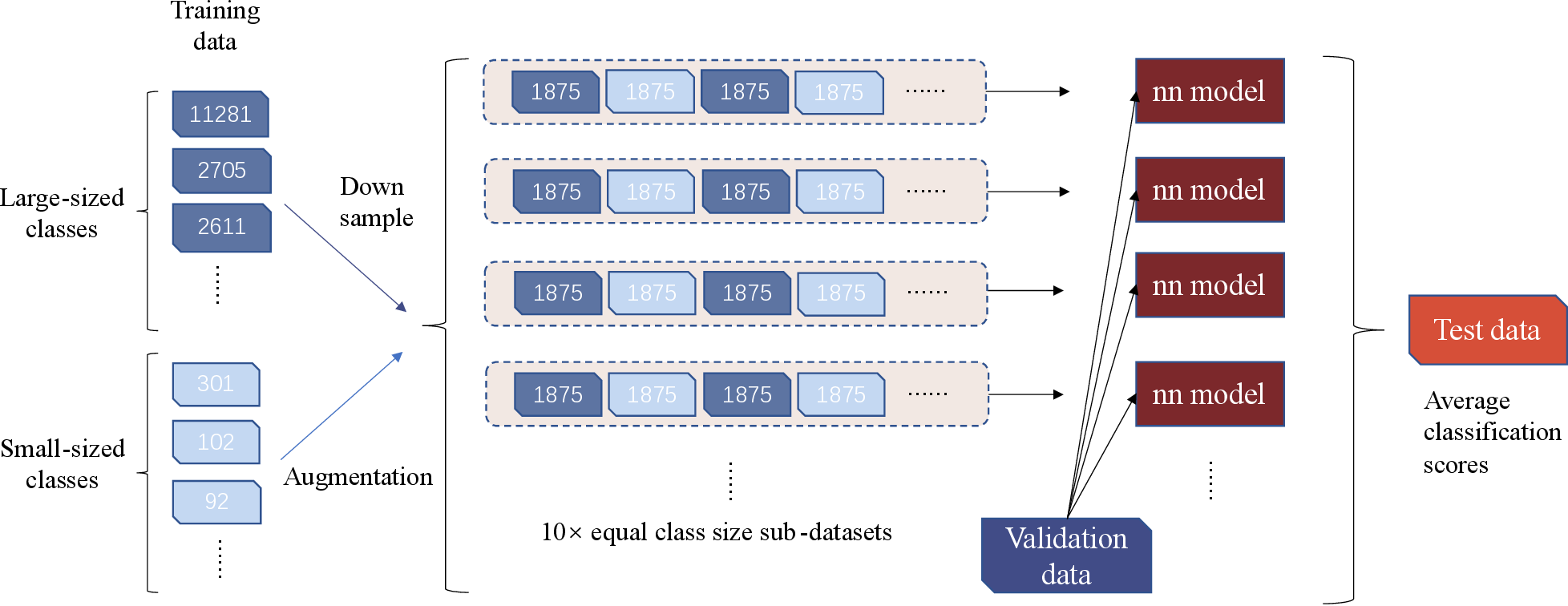}
    \caption{The augmentation-based ensemble learning process. }
    \label{augmentation}
\end{figure*}

Considering the above issues, we develop an ensemble learning method for neural networks to tackle data imbalance, as depicted in Figure~\ref{augmentation}. Like the classical ensemble manner of "bagging" \citep[][]{WOS:A1996UZ38000003}, our approach is to build several sub-datasets from the training data and then train a neural network on each sub-dataset. The classified result will be an average of all networks' outputs. When setting up sub-datasets, we apply random undersampling for large-size categories while implementing  Gaussian Process augmentation for small-size categories, ensuring every category has an equal and moderate size. Notice that the augmentation procedure generates different synthetic light curves for each sub-dataset.

Besides the benefit of handling data imbalance, the ensemble learning method has its original profit of improving performance. Our "bagging-like" approach also takes this advantage to reach higher classification accuracy at the expense of training multiple networks. However, there is no need to worry about the overall computation cost. We can train every network for only a few epochs by choosing a large learning rate and then letting the ensemble procedure combine these "weak learners" to generate a strong model.

We split the whole dataset into a training set, a validation set and a test set by a ratio of $6:1:3$ with no overlapping. The training set is used to build $10$ sub-datasets with $1875$ light curves for each class. Note that these two hyperparameters are not optimized as the computational cost is high, only a proper value to exhibit the effectiveness of ensemble operation. The validation set is augmented to have every category's size equal to the max. We do not augment the test set in order to evaluate the classification on a real data imbalance degree (actually the imbalance degree of the dataset since we do not know the natural distribution of classes). In practice, a neural network trained by this augmentation-based ensemble approach will give the same classification result for a natural light curve and its simulations.

%The two hyperparameters of sub-dataset number and sub-dataset size are not optimized, only a proper value to show effectiveness.
\section{Multi-input Neural Networks}
\label{Multi-input Neural Networks}

Neural networks have become popular in astronomical data mining with their convenience and high performance and without feature engineering, usually recurrent neural networks (RNNs) for sequential data and convolutional neural networks (CNNs) for image data. As for light curves, RNNs turned into a trendy method since \citet{2018NatAs...2..151N} demonstrated that RNNs could easily handle their characteristic of irregularly sampled time series. Meanwhile, \citet{2022ApJ...938...37S} proved that CNNs could also behave well when classifying variable stars by plotting phase-folded light curves as images. These approaches provided different choices for data with different sequence lengths. RNNs usually deal with sequences of not longer than several hundred data points, while CNNs need as many observations as possible to make the plotting have distinct patterns. We try both kinds of neural networks to classify CRTS variable stars' light curves, finding that RNN performs much better than the CNN approach because the typical sequence lengths are merely 100-300. Therefore we choose RNN as a basic structure of our neural network model. 

Apart from the light curve itself, generally, there is extra information that can be utilized as features in classification, such as period and variation amplitude (and colours in surveys other than CRTS). These numerical features demand a proper way to combine them with the RNN structure. We design a multi-input neural network to fully use these pieces of information, as shown in Figure~\ref{rnn}. Hereafter we call this network an RNN-based multi-input neural network.

We can easily combine RNN and CNN structures to perform better within this multi-input neural network architecture, as exhibited in Figure~\ref{mix}. Although using CNN alone is less effective than using RNN alone on CRTS data, the participation of CNN in the compound architecture may provide an additional perspective on high-level feature extraction. We apply a procedure of 2-stage training to optimize the compound multi-input neural network. The specification will be in Section \ref{subsec:Compound multi-input neural network}.

We deploy our neural network models on Keras, a Python deep learning API built over TensorFlow \citep[][]{tensorflow2015-whitepaper}.

\subsection{RNN-based multi-input neural network}

\begin{figure}
	\includegraphics[width=\columnwidth]{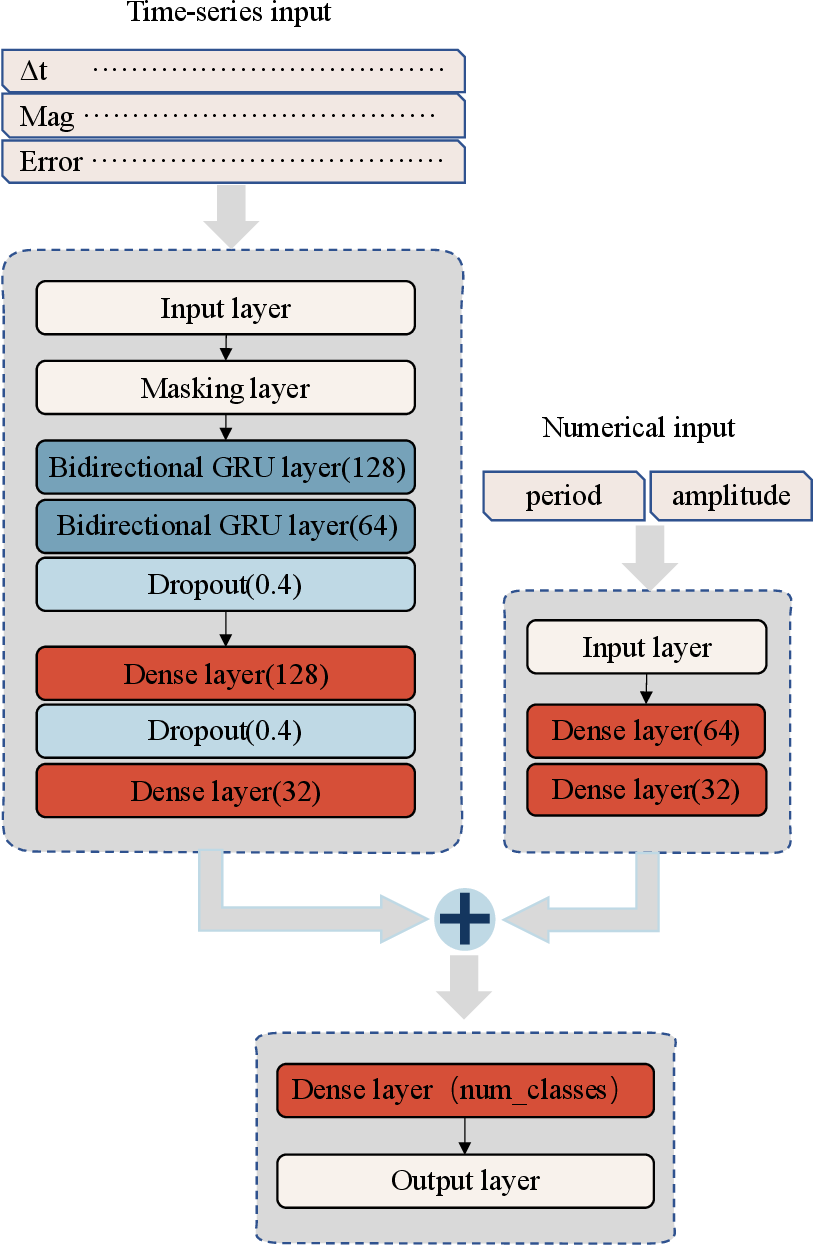}
    \caption{The architecture of the RNN-based multi-input neural network.}
    \label{rnn}
\end{figure}

As shown in Figure~\ref{rnn}, the RNN structure takes as input a sequence of vectors $(\Delta t, mag, error)$. To preprocess light curves, we transform sampling time points $t$ to time intervals $\Delta t$ and normalize the values of magnitudes to have mean zero and standard deviations one on each light curve. A masking layer is applied after the input of the sequence in order to cope with different lengths of light curves, which demand to pad input sequences with zero until they reach a same length.

The central part of this RNN structure is two bidirectional Gated recurrent unit (GRU) layers: the first returns a sequence of vectors, and the second returns an embedding vector. GRU is a widely used architecture to help retain information over a long sequence and make it possible to extract morphological features from the entire light curve. Then we employ two fully connected layers (also named dense layer) to reduce the dimensionality of the embedding vector.

We also apply two dense layers after the input layer for the numerical data to be embedded into a vector. This idea is motivated by heuristic thinking that a high-dimensional representation may attach importance to the numerical input when concatenating with high-level features of the RNN part. After the concatenation, a dense layer with softmax output gives the probability of the input belonging to each class.

To mitigate overfitting, we adopt two kinds of regularization approaches, the Dropout layer and the so-called "label smoothing" method. Dropout layers are attached to the second GRU and the first dense layers. They take effect by randomly dropping neurons and corresponding connections during training. We use a dropout rate of 0.4, which means dropping 40 per~cent of the neurons. Label smoothing is to set the one-hot label of training samples to soft labels, which is to set $1-\alpha$ for the true class and $\alpha$ for other classes, where $\alpha$ is a small number that equals 0.1 in our case. This technique can urge the neural network not to be over-confident in the classification result.

\subsection{Compound multi-input neural network}
\label{subsec:Compound multi-input neural network}
\begin{figure*}
	\includegraphics[width=\textwidth]{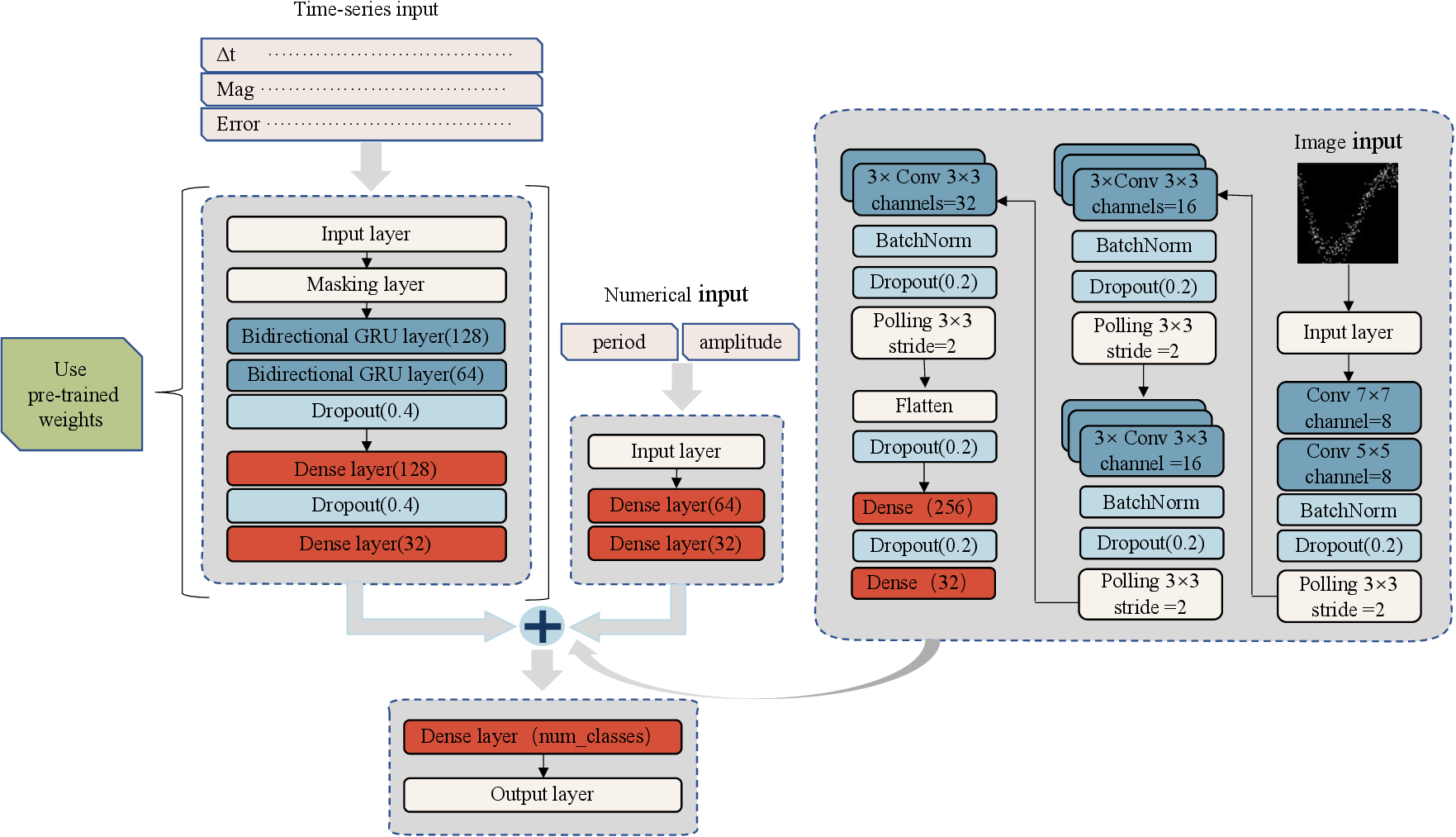}
    \caption{The architecture of the compound multi-input neural network.}
    \label{mix}
\end{figure*}

Figure~\ref{mix} shows the architecture of our compound multi-input neural network, which represents one light curve with two different inputs: a sequence and an image. The RNN and numerical input parts are identical to the previous RNN-based multi-input neural network. The image input is a $128 \times 128$ pixels single-channel image on which we plot a phase-folded light curve. The structure of the CNN part is the same as in \citet{2022ApJ...938...37S}. We only change the output of the last dense layer to a 32-dimension vector for proper concatenation with the outputs of another two parts of the network.

To train this network, we apply a 2-stage procedure. First, we ignore the image input, train an RNN-based multi-input neural network on the same dataset and save the weights of the best model. Then these weights are loaded to the compound network layer by layer, and the weights of the RNN part are fixed before training, which means setting the RNN part untrainable. This 2-stage procedure aims to make the CNN part a supplement for high-level features, not a redundant structure. The regularization approach during training for the compound network is the same as those used in the RNN-based network.

\section{implementation and results}
\label{implementation and results}
We implement our classification methods on a computer with an NVidia GeForce RTX3090 GPU. Every training epoch takes about 1 minute, and the test process takes about 3 minutes. The computation cost of the Gaussian Process augmentation is comparable to the entire training and validation process.

\subsection{Evaluation Metrics}
We adopt the confusion matrix, the balanced accuracy and the macro F1 score as metrics to evaluate the performance of the imbalanced data classification. To normalize the confusion matrix for a better exhibition, we divide each row by the total number of objects per class. Therefore the diagonal values become $Recall$ of every class. The confusion matrix can also be normalized by columns to show $Precision$ of each category. The balanced accuracy is defined as the average of $Recall$ obtained on each class, hence the average of the diagonal values of the normalized confusion matrix. The macro $F1$ score is the harmonic average of $Precision$ and $Recall$, as follow

\begin{equation}
    F1 =2\times \frac{ Precision\times Recall}{ Precision+Recall}
\end{equation}
The higher the macro F1 score, the better the classification result.

\subsection{Hyperparameter setting}
Most of the hyperparameters are depicted in Figures~3-4, leaving the batch size and the learning rate for the Adam optimizer. These two hyperparameters are always tuned together. We choose a relatively small batch size of 32 to get a better generalization performance for the neural network model and a relatively large learning rate of 0.001 to reduce the computation cost. We also try lower learning rates and careful scheduling strategies, finding the improvement level less than the effect of different random initial weights of neural networks. Considering the ability of the ensemble learning method to integrate weak learners into a strong learner,  it is rational to make this trade-off between performance and computation cost.

\subsection{Results of the RNN-based multi-input neural network}
\begin{figure}
	\includegraphics[width=0.84\columnwidth]{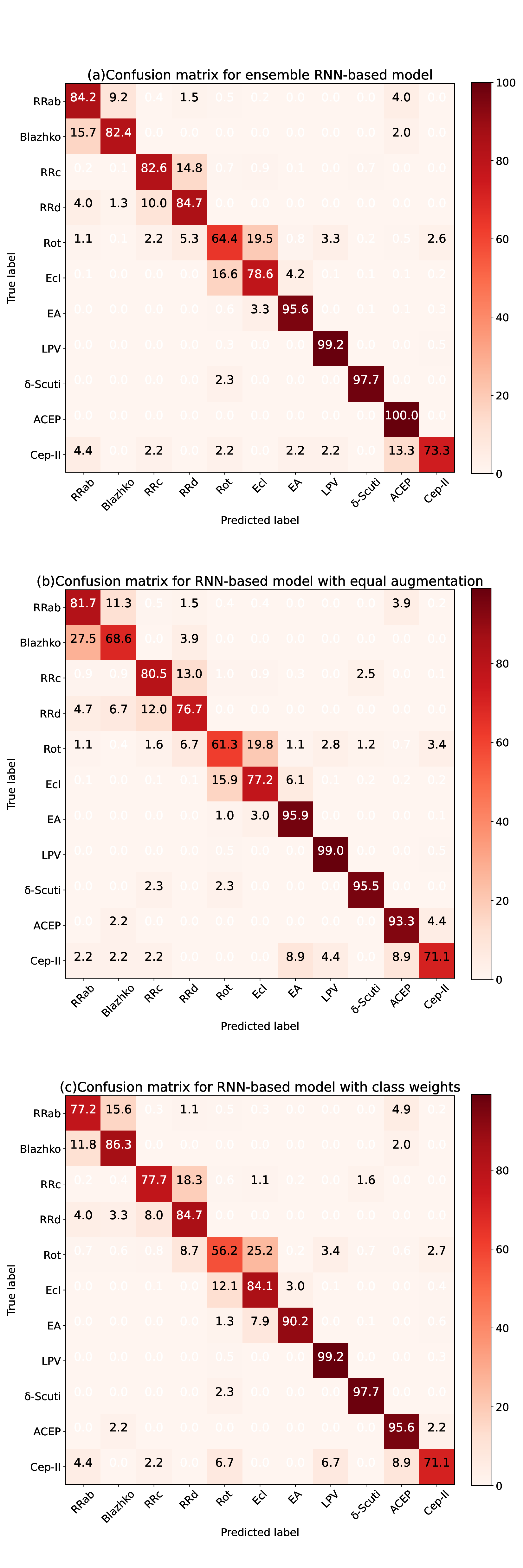}
    \caption{The confusion matrixes of the ensemble RNN-based model with comparison to the other two approaches. The matrixes are normalized to the $Recall$ view. (a): classification result of the ensemble RNN-based model. (b): the result of the equal augmentation approach. (c): the result of moderate augmentation and class weights.}
    \label{fig:RNN_result}
\end{figure}

The training step of the RNN-based model takes ten epochs on each sub-dataset, therefore 100 epochs for the entire ensemble learning process. Figure~\ref{fig:RNN_result} shows the confusion matrix of the classification results on the test data compared to the other two methods: one is to apply equal augmentation, i.e. augmenting each category of the training set to the same number of the largest category; another is to augment the small-sized classes to a moderate size (1875 in the implementation) and employ the class weight when training. The macro F1 score of the ensemble technique, the equal augmentation, and augmentation with class weights are $0.71$, $0.66$, $0.67$, respectively. As a result, the performance with ensemble technique is obviously superior to that with the equal augmentation and augmentation with class weights.

\subsection{Result of the compound multi-input neural network}
\begin{figure}
	\includegraphics[width=0.9\columnwidth]{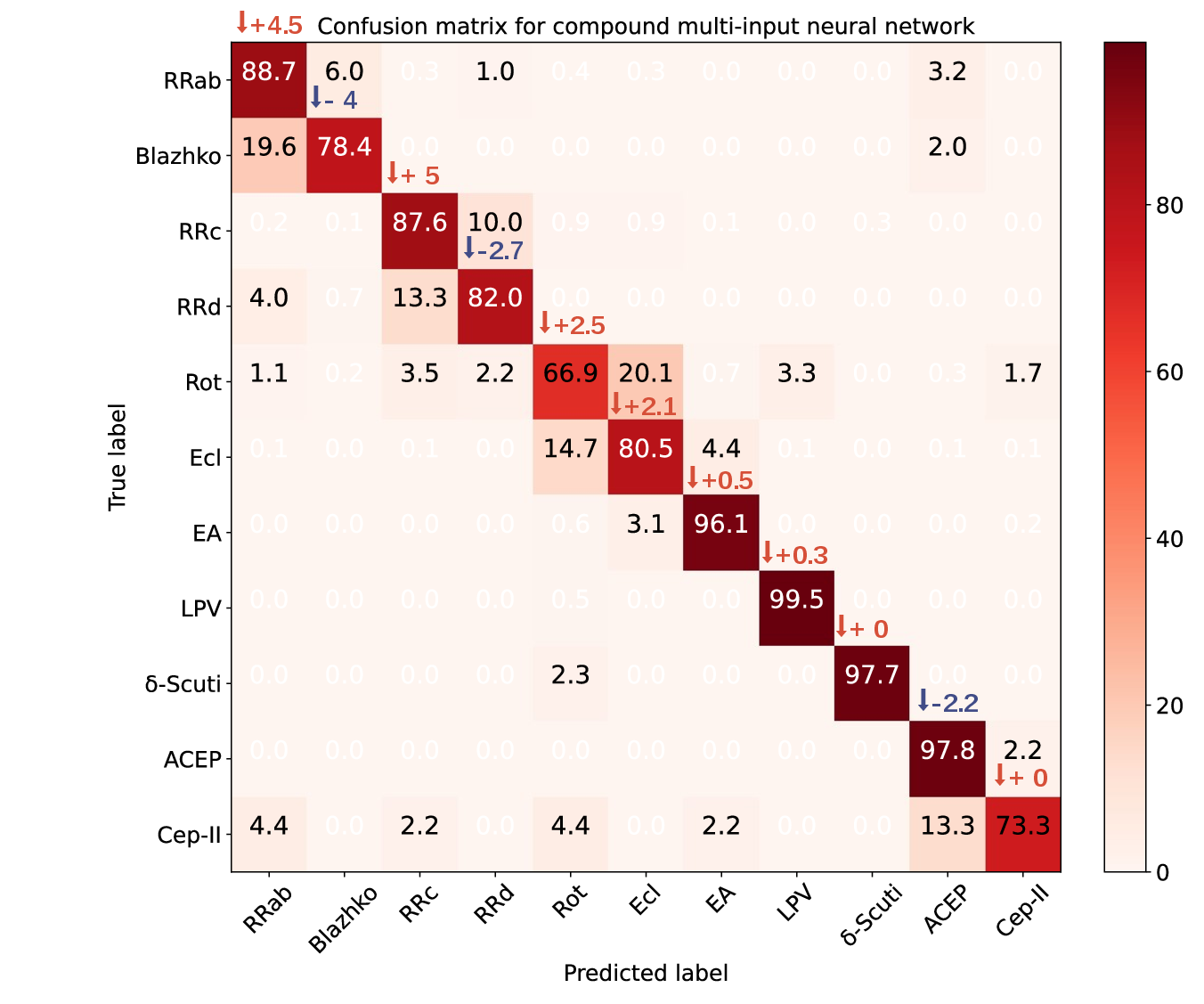}
    \caption{The confusion matrix of the ensemble compound multi-input neural network, marked with the variation of each class's $Recall$ compared to the RNN-based network. }
    \label{after_image}
\end{figure}

Though training a single RNN-based or CNN-based multi-input neural network consumes not much time, the compound neural network is hard to train by taking 100 epochs on every sub-dataset to find the optimal model. Table~\ref{tab:evaluation} exhibits the balanced accuracy of the results given by our two different networks trained on each sub-datasets. Both the classification score on validation and test sets are offered. The final ensemble learning result of the compound network is displayed in Figure~\ref{after_image}, where we mark the variation of each class's $Recall$ compared to the RNN-based network. 

\begin{figure*}
	\includegraphics[width=\textwidth]{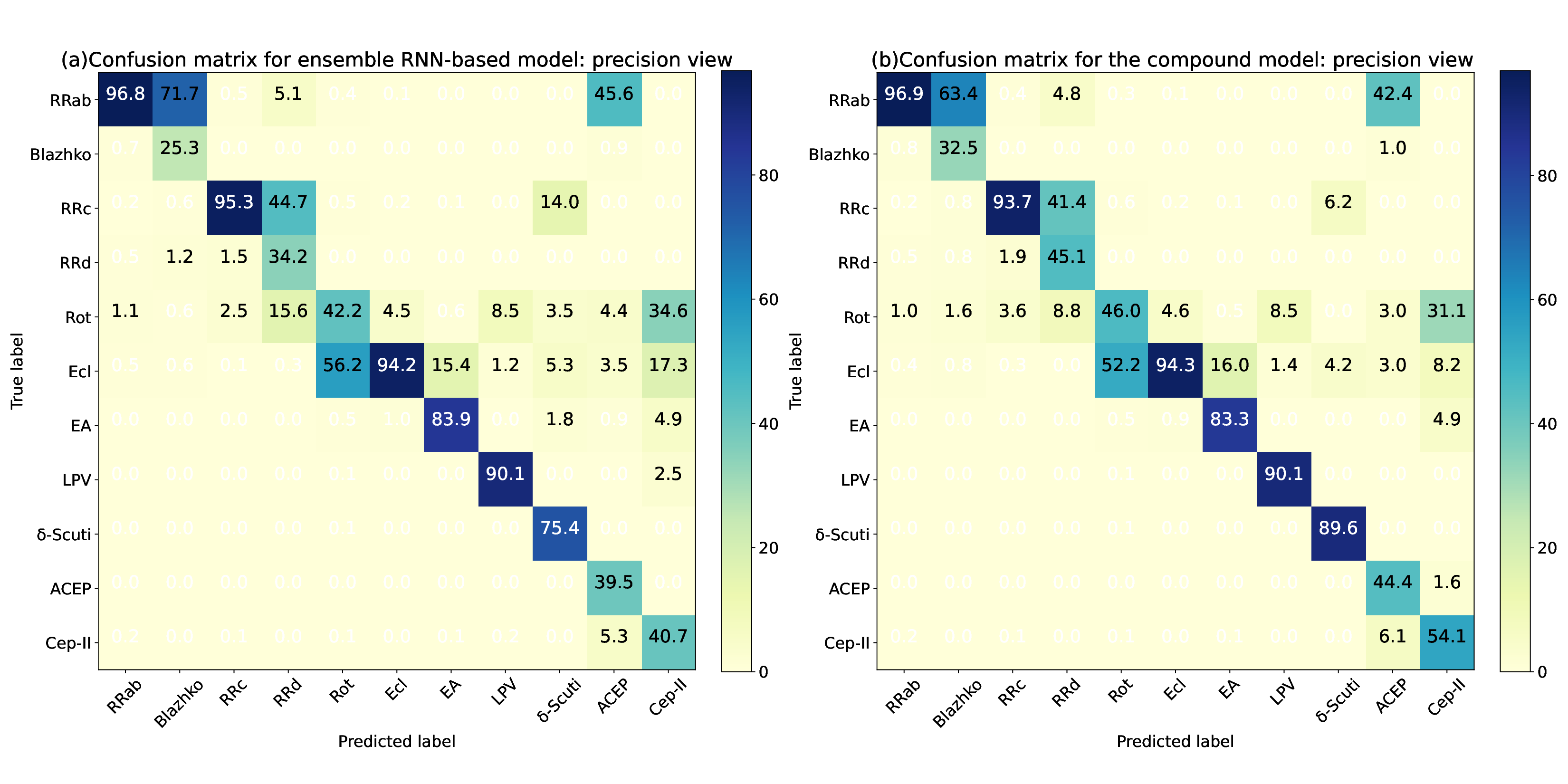}
    \caption{The confusion matrixes normalized to the $Precision$ view. (a): the result of the RNN-based model. (b): the result of the compound model. On the imbalance test set, the misclassified samples of the large-sized classes can easily dominate samples of the predicted small-sized categories.}
    \label{precision_2cm}
\end{figure*}

The overall balanced accuracy of the compound model improves slightly from $85.7$ per~cent to $86.2$ per~cent compared to the RNN-based model, while the macro $F1$ score increases from 0.71 to 0.75. The relatively significant improvement in the macro $F1$ score derives from the advance in $Precision$, which is depicted in Figure~\ref{precision_2cm} by the confusion matrix normalized to the $Precision$ view. The compound model classifies the large-sized categories more accurately at the cost of slightly decreasing $Recall$ of the small-sized ones, and this relieves the misclassification of the majority of the samples.  

\begin{table}
	\centering
	\caption{The balanced accuracy of models trained on each of the ten sub-datasets, given by two different neural networks of the RNN-based and compound networks. Both the classification score on validation and test sets are offered.}
	\label{tab:evaluation}
	\begin{tabular}{llll} % four columns, alignment for each
		\hline
		RNN val & Compound val & RNN test & Compound test \\
        (\%)    &(\%)          &(\%)      &(\%)\\
  \hline
        80.91 & \textbf{84.29} & 81.90 & \textbf{84.97}\\
        82.68 & \textbf{83.49} & 82.62 & \textbf{84.29}\\
        82.19 & \textbf{84.62} & 83.43 & \textbf{85.10}\\
        81.90 & \textbf{82.19} & \textbf{82.78} & 79.92\\
        82.82 & \textbf{83.54} & 82.67 & \textbf{82.72}\\
        81.57 & \textbf{83.69} & 80.74 & \textbf{81.25}\\
        83.00 & \textbf{85.33} & 83.92 & \textbf{85.34}\\
        84.61 & \textbf{84.97} & 83.39 & \textbf{84.77}\\
        80.74 & \textbf{83.24} & 78.62 & \textbf{84.30}\\
        83.13 & \textbf{83.20} & \textbf{79.73} & 78.95\\
		\hline
	\end{tabular}
\end{table}

\subsection{10-fold cross-test}
The performance of a classification algorithm on a dataset usually varies with different partitions of training and test set. To reliably demonstrate the capacity of our method, we carry out a 10-fold cross-test. We split the original dataset into ten equal-sized distinct parts with no overlapping. Each is used as a test set in a training-and-evaluation process, while others are for training and validation sets. Thus the process is repeated ten times. The augmentation and ensemble learning operations are implemented respectively on each of the ten processes.

\begin{figure*}
	\includegraphics[width=\textwidth]{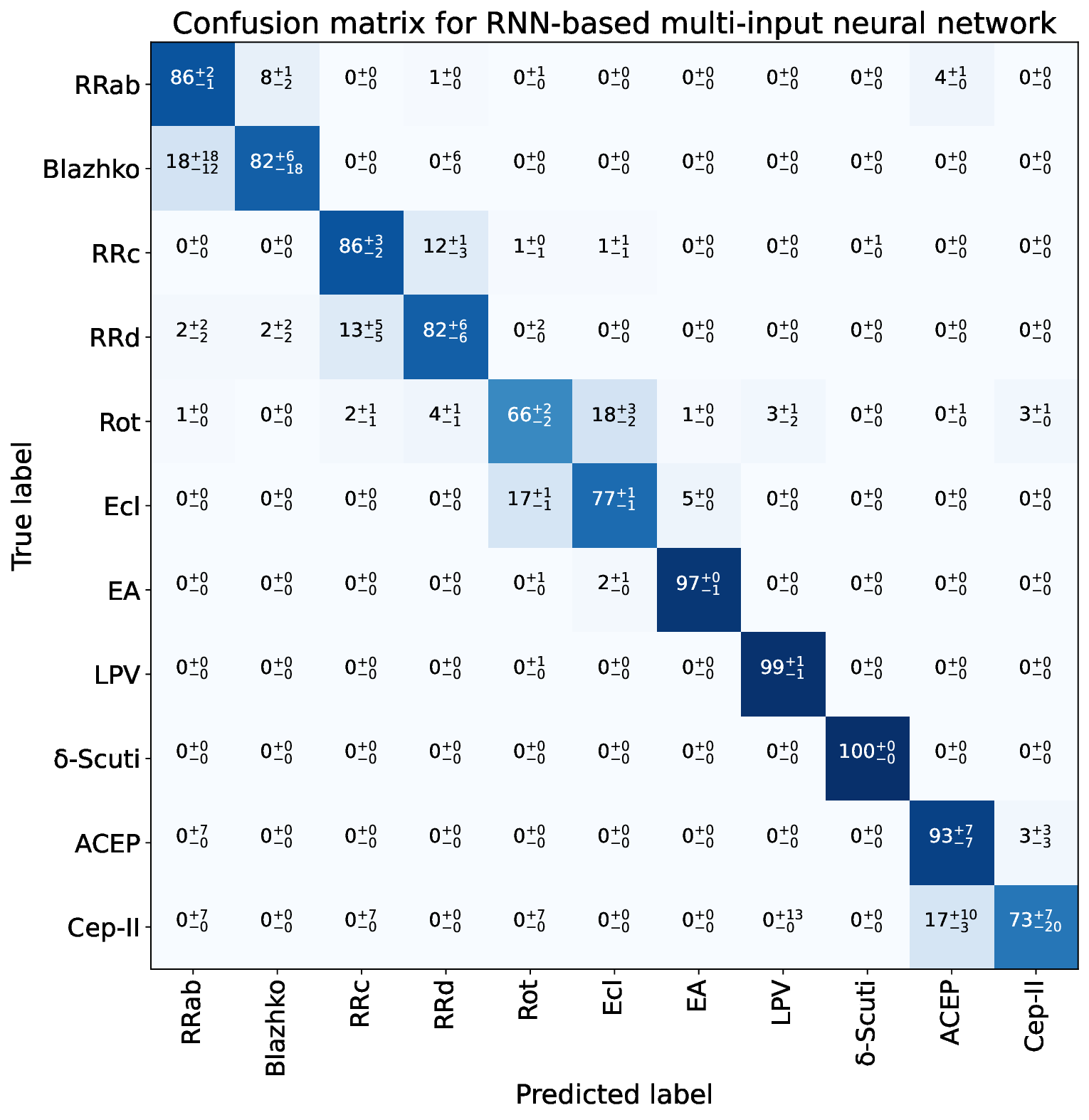}
    \caption{The 10-fold cross-test result for the RNN-based multi-input neural network. The matrix is normalized to the $Recall$ view. Each matrix element is the median of the corresponding values in the ten confusion matrixes. The superscript and the subscript indicate the deviation of the second-best and second-worst values, respectively.}
    \label{cross_test}
\end{figure*}

Here we only perform cross-test for the RNN-based multi-input neural network since the compound model only slightly improves on $Recall$. Figure~\ref{cross_test} depicts the result. Each matrix element is the median of the corresponding value in the ten confusion matrix. The superscript and the subscript indicate the deviation of the second-best and second-worst values, respectively. On the imbalance test set, the relatively large deviation for small-size classes can derive from the misclassification of only one or two samples.

\section{Discussion}
\label{sec:discussion}

Considering all types of RR Lyrae stars as a whole, the compound neural network achieves a total $Recall$ of $96.7$ per~cent and a total $Precision$ of $97.1$ per~cent. Other periodic variable stars, except rotational stars, rarely contaminate the RR Lyrae star samples. However, some normal RRab stars are misclassified as RR Lyrae stars with the Blazhko effect or ACEP stars. Due to the significant data imbalance, these misclassified samples constitute a large proportion of the predicted samples for these two small-sized categories. Similarly, the misclassified samples from RRc stars greatly affect the purity of RRd stars.

For rotational stars, the compound model achieves $66.9$ per~cent of $Recall$ and $42.2$ per~cent of $Precision$. Some of these stars could confuse with the contact and semi-detached binary (Ecl) stars. Moreover, the misclassified rotating stars markedly contaminated the small-sized Cep-\uppercase\expandafter{\romannumeral2} stars class. The Ecl class gets $Recall$ of $80.5$ per~cent and $Precision$ of $94.2$ per~cent. The misclassified samples of Ecl stars also contaminate the detached binary stars (EA) because there is no clear division between these two types. $Recall$ and $Precision$ of EA stars are $96.1$ per~cent and $83.3$ per~cent, respectively.

The long period variable stars (LPV) and the $\delta$-Scuti stars are relatively distinct from other types of stars, and have $Recall$ of $99.5$ per~cent and $97.7$ per~cent, respectively. But they could be polluted by the misclassified samples of other large-sized classes, which makes $Precision$ decrease to $90.1$ per~cent and $89.6$ per~cent, for LPV and $\delta$-Scuti respectively.

For both types of cepheids, the total $Recall$ is $93.3$ per~cent while the total $Precision$ is $52.5$ per~cent. The ACEP stars have $Recall$ of $97.8$ per~cent and while Cep-\uppercase\expandafter{\romannumeral2} stars have $73.3$ per~cent. They mainly confuse each other, but also get dramatically contaminated by other large-sized classes' misclassified samples.

%In Section \ref{subsec:Generate Synthetic Light Curves}, for simplicity, we scale up the posterior uncertainty $\pmb{\sigma}_*$ to let the synthetic light curves have the same uncertainty degree as their prototypes. This operation aims at more `real' augmentation data but not in the most appropriate way. The GP actually build a model for the intrinsic uncertainty of the data, which is always lower at sections with enough data and higher at those with little data. The simple scale-up operation will underestimate the error of synthetic light curves at data-rich sections while overestimating at data-poor sections, with the photometric error not considered. A better approach would be to add a well-modelled photometric error to the GP posterior uncertainty and tune the result to a proper uncertainty degree. 

\section{Conclusions}
\label{Conclusions}
In this paper, we present an ensemble learning approach based on data augmentation for periodic variable star classification by light curves using deep learning on imbalanced data. We apply Gaussian Process to generate artificial light curves with uncertainties for small-size classes and take undersampling for large-size classes, setting up balanced sub-datasets of the training set. Training models on these sub-datasets could avoid overfitting on small-size classes, and the ensemble result shows performance improvement.

We design two kinds of neural network architectures for the task: the RNN-based multi-input model and the compound model combing RNN and CNN structures. These multi-input models take both the light curve and additional numerical features as input. On the CRTS variable star data, The macro $F1$ score on the imbalanced test set reaches $0.71$ and $0.75$ for the RNN-based and compound model, respectively. 

Our ensemble learning approach can easily cooperate with different deep learning models since it is a data-level technique. Our attempt to combine CNN and RNN structures implies that using different representations of light curves together in a model is possible for higher performance. This kind of compound neural network architecture is flexible for time series sky surveys with different light curve lengths. These methods put forward by us will contribute to a better classification of variable sources with time series data in the future projects (e.g. LSST), finish the multi-classification in one step with high performance, and also shed light on the imbalance classification with multimodal data.

\section*{Acknowledgements}
This paper is funded by the National Natural Science Foundation of China (Grant Nos.12273076, 12203077, 12133001, U1831126 and 11873066), the Science Research Grants from the China Manned Space Project (Nos. CMS-CSST-2021-A04 and CMS-CSST-2021-A06), and Natural Science Foundation of Hebei Province (No.A2018106014). 

\bibliography{literatures}{}
\bibliographystyle{aasjournal}

\end{document}